\documentclass[aps,pra,twocolumn,groupedaddress,showpacs]{revtex4}
\usepackage{amsmath}
\usepackage{graphicx}% Include figure files
\usepackage{dcolumn}% Align table columns on decimal point
\usepackage{bm}% bold math

\begin{document}

% Use the \preprint command to place your local institutional report
% number in the upper righthand corner of the title page in preprint mode.
% Multiple \preprint commands are allowed.
% Use the 'preprintnumbers' class option to override journal defaults
% to display numbers if necessary
%\preprint{}

\title{Bound-bound transitions in the emission spectra of Ba$^{+}$--He excimer}

% repeat the \author .. \affiliation  etc. as needed
% \email, \thanks, \homepage, \altaffiliation all apply to the current
% author. Explanatory text should go in the []'s, actual e-mail
% address or url should go in the {}'s for \email and \homepage.
% Please use the appropriate macro foreach each type of information

% \affiliation command applies to all authors since the last
% \affiliation command. The \affiliation command should follow the
% other information
% \affiliation can be followed by \email, \homepage, \thanks as well.

\author{P. Moroshkin$^{1}$}
\email[]{petr.moroshkin@riken.jp}
\author{K. Kono$^{1,2,3}$}

\affiliation{$^{1}$RIKEN, CEMS, 2-1 Hirosawa, Wako, 351-0198 Saitama, Japan}

\homepage[]{http://www.riken.jp/en/research/labs/cems/qtm_inf_electron/qtm_condens_phases/}

\affiliation{$^{2}$Institute of Physics, NCTU, Hsinchu 300, Taiwan\\ $^{3}$Institute of Physics, KFU, Kazan, Russia}

\date{\today}

\begin{abstract}

We present an experimental and theoretical study of the emission and absorption spectra of the Ba$^{+}$ ions and Ba$^{+\ast}$He excimer quasimolecules in the cryogenic Ba--He plasma. We observe several new spectral features in the emission spectrum which we assign to the electronic transitions between bound states of the excimer correlating to the 6$^{2}P_{3/2}$ and 5$^{2}D_{3/2,5/2}$ states of Ba$^{+}$. The resulting Ba$^{+}$(5$^{2}D_{J}$)He is a metastable electronically excited complex with orbital angular momentum $L$=2, thus expanding the family of known metal--helium quasimolecules. It might be suitable for high-resolution spectroscopic studies and for the search for new polyatomic exciplex structures.

\end{abstract}

% insert suggested PACS numbers in braces on next line
\pacs{33.70.-w, 33.50.-j, 32.50.+d}

\maketitle

% body of paper here - Use proper section commands
% References should be done using the \cite, \ref, and \label commands

\section{Introduction \label{Seq:Introduction}}

Helium does not form any chemical compounds.
An overlap between the filled $1s$ orbital of He and the electron density of any other atom leads to strong repulsion according to the Pauli principle.
However, quasimolecules (bound states) formed by one or several He atoms and an electronically excited metal atom are well known and are referred to as excimers and exciplexes.
In these complexes, He atoms are arranged around a highly anisotropic valence electron orbital of the excited metal atom in such a way that there is no overlap of the electronic clouds.
Pauli repulsion therefore does not play a role and the complex is bound by van der Waals forces.
These quasimolecules play an important role as model systems in quantum chemistry and in the physics of quantum fluids \cite{JakubekCPL1997,NakayamaJCP2001,TakayanagiPCCP2004,LeinoJCP2008,DellAngeloJCP2012,ChattopadhyayJPB2012,CargnoniPCCP2013}.
They have been observed experimentally in cold He gas \cite{EnomotoPRA2002,HiranoPRA2003,FukuyamaPRA2004}, bulk liquid He \cite{PerssonPRL1996,HiranoPRA2003}, and solid \cite{NettelsPRL2005,MoroshkinJCP2006,HoferPRA2006} He, and on He nanodroplets \cite{RehoFD1997,BruhlJCP2001,SchulzPRL2001,GieseJCP2012}.
There have also been related studies on excimer molecules formed by heavier rare gases (reviewed in \cite{GallagherTAP1984}) and a more recent work on long-range metal--helium van der Waals complexes formed at ultralow temperatures in buffer gas traps \cite{BrahmsPRL2010,BrahmsPCCP2011}.

Most known metal--helium complexes have electronic orbital angular momentum $L$ = 1, which corresponds to a $P$ state of the metal atom.
They are observed and identified via their spontaneous emission transitions terminating at the unbound electronic ground state.
Owing to very strong Pauli repulsion in the ground state, all these bound-free transitions are spectrally very broad and strongly red-shifted with respect to the corresponding transitions of a free metal atom.

The excimer formed by an electronically excited Ba$^{+}$ ion in the $L$ = 1 state and a ground-state He atom has been previously studied both theoretically \cite{MellaJPCA2014,LealJCP2016} and experimentally \cite{FukuyamaPRA2004,FukuyamaPRA2007}.
Here we report an experimental observation of the Ba$^{+\ast}$He complex in an electronic state with a higher orbital angular momentum, $L$ = 2, \textit{i.e.}, a $D$-state.
We obtain a series of electronic transitions between the quasimolecular states correlating to the 5$D$ and 6$P$ states of Ba$^{+}$.
For these transitions, both the initial and the final state are bound states.
Therefore, they are spectrally sharp and reveal a well-resolved vibrational structure.
To the best of our knowledge, transitions of this type (bound-bound) are observed in the emission spectra of metal--He excimers and exciplexes for the first time.
The closest analogues are the visible and near-infrared bands of the He$_{2}^{\ast}$ excimer \cite{DennisPRL1969,MendozaLunaEPJD2013}.

\section{Experiment \label{Seq:Experiment}}

\begin{figure} [tbp]
    \includegraphics [width=\columnwidth] {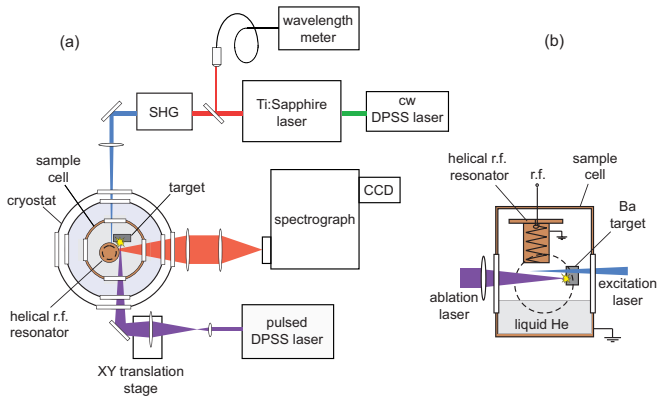}
    \caption{(a) Top view of the experimental setup, (b) vertical cross section of the sample cell.}
    \label{fig:Setup}
\end{figure}

The experiments were performed in an optical He bath cryostat cooled
to $T$ = 1.35--1.7 K by pumping on the He bath.
A top view of the cryostat and the optical setup is shown in Fig.
\ref{fig:Setup}(a) and a vertical cross section of the sample cell is shown in
Fig. \ref{fig:Setup}(b).
The cell was half-filled with liquid He.
Ba atoms and clusters were produced by laser ablation from a Ba metal
target positioned above the liquid He.
The ablation was performed by a frequency-tripled pulsed diode-pumped solid-state (DPSS) laser ($\lambda$ = 355 nm) operating at a repetition rate of 100 Hz with a pulse energy of 70 $\mu$J.
The ablation laser beam enters the sample cell through a side window.
The expanded laser beam is focused by an $f$ = 10 cm lens mounted on
an XY translation stage.
During the laser ablation, the lens is constantly moving in a plane
orthogonal to the optical axis.
The ablation spot is thus moving along the target surface and the
laser pulse always hits a fresh spot.
In this way, we reduce the hole-drilling effect and improve the
stability of the ion yield.

The ablation products are ionized  by the radio frequency (r.f.)
discharge created inside an r.f. helical resonator with an open
lower end.
The resonator is driven at its resonance frequency of 433 MHz.
The output power of the r.f. amplifier is varied in the range of 1--10 W, which results in heating of the sample cell by 10--100 mK.

Ba$^{+}$ ions in the He gas above the liquid He are excited by a cw
laser beam of a frequency-doubled tunable Ti:sapphire laser pumped
by a cw green DPSS laser ($\lambda$ = 532 nm).
The cw laser beam crosses the sample cell in the opposite direction to the ablation laser.
The output wavelength of the Ti:sapphire laser is controlled by a wavelength meter with an absolute accuracy of 200 MHz.
The laser is tuned to $\lambda$ = 911.0617 nm (10976.205 cm$^{-1}$).
At this wavelength, the output radiation of the second harmonic
generator (SHG) is in resonance with the 6$^{2}S_{1/2} \rightarrow$
6$^{2}P_{3/2}$ ($D_{2}$) absorption line of Ba$^{+}$ in vacuum.

The laser-induced fluorescence of Ba$^{+}$ together with the
emission of the r.f. plasma at the open end of the helical resonator
is collected at 90$^{\circ}$ with respect to both laser beams and is
imaged on the entrance slit of a grating spectrometer equipped
with a CCD camera.

\begin{figure} [tbp]
    \includegraphics [width=\columnwidth] {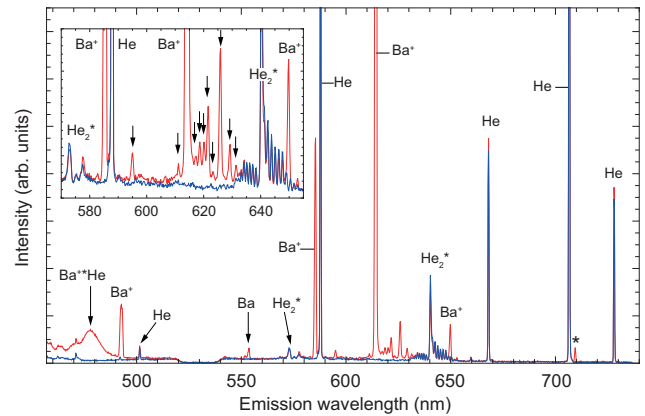}
    \caption{Spectra of laser-induced fluorescence and r.f. plasma emission.
    Blue line: ablation off, red line: ablation on. $T$ = 1.47 K, $p_{He}$=4 mbar.
    Positions of new spectral lines are indicated in the inset by vertical arrows. The peak marked with an asterisk is an artifact.}
    \label{fig:SpecOverview}
\end{figure}

A typical pair of spectra recorded under the same conditions, but with and without laser ablation is shown in Fig. \ref{fig:SpecOverview}.
With the ablation laser turned off, the observed spectrum consists
of spectral lines of He atoms and He$_{2}^{\ast}$ excimers produced
in the r.f. plasma.
With the ablation laser switched on, the spectrum is dominated by the lines of Ba$^{+}$.
We verified that this is indeed laser-induced fluorescence
and not the emission by Ba$^{+}$ ions excited by collisions in the
r.f. plasma or in the ablation plume.
The excitation spectrum of this emission obtained by scanning the Ti:sapphire laser and recording the fluorescence intensity of the strongest line at 614.2 nm is shown in Fig. \ref{fig:SpectraTheorExp}(a).
The lineshape is fitted with Lorenzian profile shown by the solid
line.
The center wavelength coincides within the accuracy of the
measurement with that of the 6$^{2}S_{1/2} \rightarrow 6^{2}P_{3/2}$ transition in a free ion and the FWHM spectral width is
equal to 1.79$\pm$0.07 GHz.
Under these conditions, the spectral lines of neutral Ba atoms excited by the r.f. plasma are much weaker.
Only the strongest line of the Ba atoms can be distinguished in Fig. \ref{fig:SpecOverview}, which is the $6s6p ^{1}P_{1} \rightarrow 6s^{2}$  $^{1}S_{2}$ transition at 553.5 nm.

\begin{figure} [tbp]
    \includegraphics [width=\columnwidth] {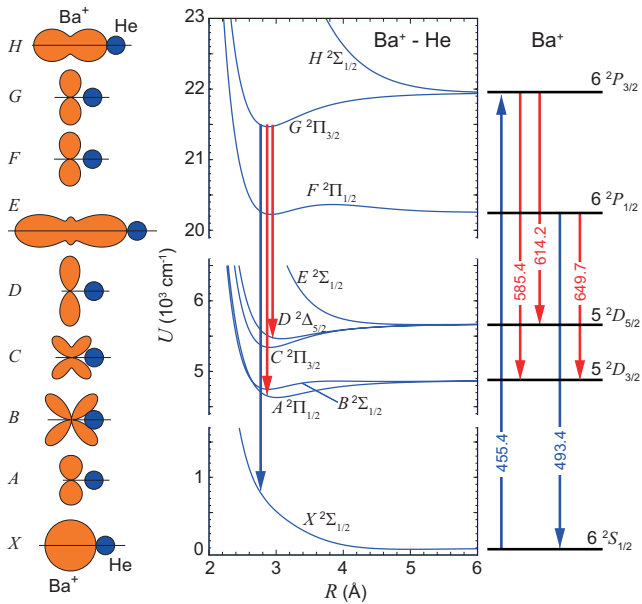}
    \caption{Energy levels of a free Ba$^{+}$ ion (right side) and the Ba$^{+}$--He quasimolecule \cite{MellaJPCA2014} (center). On the left, electron orbitals of Ba$^{+}$ and He are drawn schematically. All orbitals are symmetric with respect to rotations around the horizontal axis.}
    \label{fig:PotentialDiagram}
\end{figure}

The observed transitions of Ba$^{+}$ and their wavelengths are shown
on the right-hand side of Fig. \ref{fig:PotentialDiagram}.
The laser excitation populates the 6$^{2}P_{3/2}$ state at 21953
cm$^{-1}$, which has three channels for radiative decay (via
spontaneous emission): to the ground state 6$^{2}S_{1/2}$ and to the
two lower-lying states 5$^{2}D_{3/2}$ and 5$^{2}D_{5/2}$ with
wavelengths of 455.4, 585.4, and 614.2, respectively.
Another two prominent spectral lines in Fig. \ref{fig:SpecOverview}
correspond to the transitions from the 6$^{2}P_{1/2}$ state to the
ground state 6$^{2}S_{1/2}$ and to the 5$^{2}D_{3/2}$ state at 493.4
and 649.7 nm, respectively.
The 6$^{2}P_{1/2}$ state lies 1700 cm$^{-1}$ below the laser-excited
6$^{2}P_{3/2}$ state and can be populated by fine-structure mixing
atomic collisions in the plasma.

The broad spectral feature at 460-480 nm is assigned to the emission
of the Ba$^{+\ast}$He excimer \cite{FukuyamaPRA2004,FukuyamaPRA2007}.
The electronic energy levels of this complex are shown in the central part of Fig. \ref{fig:PotentialDiagram}.
The quasimolecule is formed upon the laser excitation of the
Ba$^{+}$ to the 6$^{2}P_{3/2}$ state along the attractive
potential curve labeled $G ^{2}\Pi_{3/2}$, and it decays via a spontaneous emission to the $X ^{2}\Sigma_{1/2}$ ground state.

Small variations of the cell temperature in the range of 1.4 - 1.7 K lead to large changes in the intensities of the two lines originating from
the 6$^{2}P_{1/2}$ state and the excimer emission.
Typically, the intensity increases with the temperature, although the fluctuations are too large for a detailed study.
As the density of the He vapor rapidly increases with the
temperature, the probability of collisions with He atoms increases
and the population of these states is expected to grow.

Several smaller spectral features exist in the range of 590 - 630 nm that appear
together with the spectral lines of Ba$^{+}$ and the Ba$^{+\ast}$He
excimer.
In the inset of Fig. \ref{fig:SpecOverview} we show a magnification of the
relevant part of the spectrum with the positions of the new spectral
lines marked by vertical arrows.
These lines cannot be identified with any transitions of Ba$^{+}$,
or neutral Ba and He atoms, neither with transitions of Sr or Ca,
which can be contained in the ablation target as impurities.
They are excited only with the Ti:sapphire laser tuned exactly
to the wavelength of the 6$^{2}S_{1/2} \rightarrow 6 ^{2}P_{3/2}$ absorption line,
in the same way as for all other spectral lines discussed above.
The temperature dependence of their intensity is similar to that of Ba$^{+\ast}$He excimers and the transitions originating from the 6$^{2}P_{1/2}$ state.

\section{Theoretical model \label{Seq:Theory}}

We interpret the newly discovered spectral lines as a band of
transitions between the electronic states of the Ba$^{+\ast}$He
complex (excimer) correlating to the 6$^{2}P_{3/2}$ and 5$^{2}D_{3/2,5/2}$ electronic states of Ba$^{+}$.
Our theoretical model is based on the set of \textit{ab initio} Ba$^{+}$--He pair potentials calculated in \cite{MellaJPCA2014}.
These potential curves depend on the orientation of the orbital
angular momentum of the valence electron of Ba$^{+}$ with respect to
the internuclear axis but do not take into account the spin-orbit
coupling.
The spin-orbit coupling is included in the model following the
standard approach.
We write the total Hamiltonian of the complex $H_{tot}(R)$ as the sum
of the spin-orbit term $H_{SO}$ and the interaction Hamiltonian
$H_{int}(R)$, which is diagonal in the $|L,M_{L},S,M_{S}\rangle$ basis.
Here, $R$ denotes the interatomic distance, $L$ and $S$ are the orbital angular momentum and the spin of the valence electron of Ba$^{+}$, and $M_{L}$ and $M_{S}$ are the projections of $L$ and $S$ onto the internuclear axis, respectively.
The adiabatic potential curves are obtained by the numerical
diagonalization of $H_{tot}(R)$ and are shown in the central part
of Fig. \ref{fig:PotentialDiagram}.
The resulting eigenstates are $R$-dependent linear combinations of the  $|L,M_{L},S,M_{S}\rangle$ states of Ba$^{+}$.
On the left-hand side of Fig. \ref{fig:PotentialDiagram} we show angular distributions of the valence electron density, corresponding to these states.

The excimer is formed in the potential well of the $G ^{2}\Pi_{3/2}$ state correlating to the 6$^{2}P_{3/2}$ asymptote.
It decays via spontaneous emission towards the lower-lying states corellating to the 6$^{2}S_{1/2}$, 5$^{2}D_{3/2}$, and 5$^{2}D_{5/2}$ asymptotes.
First, we numerically solve the single-dimensional Schr\"{o}dinger
equation and find the eigenenergies $E(v)$ and wavefunctions
$\Psi(v,R)$ of all bound vibration states in the potential wells of
the states $A ^{2}\Pi_{1/2}$, $B ^{2}\Sigma_{1/2}$, $C
^{2}\Pi_{3/2}$, $D ^{2}\Delta_{5/2}$, and $G ^{2}\Pi_{3/2}$.
Their energies are given in Table \ref{table:Energies}.

\begin{table}
    \centering
    \begin{tabular}{c|c|c|c|c|c|c}
        $state$ & $U_{well}$ & $v$ = 0 & $v$ = 1 & $v$ = 2 & $v$ = 3 & $v$ = 4 \\\hline\hline
        $A ^{2}\Pi_{1/2}$ & 244 & 55 & 141 & 196 & 226 & -- \\
        $B ^{2}\Sigma_{1/2}$ & 127 & 61 & 116 & -- & -- & -- \\
        $C ^{2}\Pi_{3/2}$ & 335 & 70 & 183 & 261 & 306 & -- \\
        $D ^{2}\Delta_{5/2}$ & 212 & 51 & 127 & 175 & -- & -- \\
        $G ^{2}\Pi_{3/2}$ & 485 & 80 & 216 & 321 & 396 & 444 \\
    \end{tabular}
    \caption{Well depths $U_{well}$ and eigenenergies of vibration states of Ba$^{+\ast}$He excimer (in cm$^{-1}$).}\label{table:Energies}
\end{table}

\begin{figure} [tbp]
	\includegraphics [width=\columnwidth] {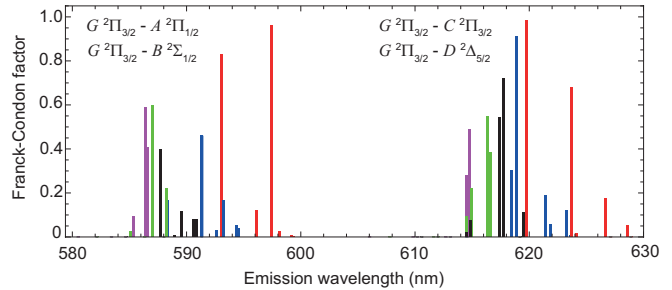}
	\caption{Calculated Franck-Condon factors for the transitions from $G ^{2}\Pi_{3/2}$ state of Ba$^{+\ast}$He towards
		$A ^{2}\Pi_{1/2}$, $B ^{2}\Sigma_{1/2}$, $C ^{2}\Pi_{3/2}$, and $D ^{2}\Delta_{5/2}$ states. Vertical bars colored in red, blue, black, green, and magenta
		correspond to the transitions originating from the states with $v$ = 0, 1, 2, 3, and 4, respectively.}
	\label{fig:CondonFactors}
\end{figure}

Then we compute the Franck-Condon factors for the transitions between the vibration states belonging to different electronic states.
\begin{equation}
F (v,v') = \left( \int \Psi_{e}(v,R) \Psi_{a}^{\ast}(v',R) dR \right)^{2}  \label{eq:FranckCondon}
\end{equation}
Here, $\Psi_{e}(v,R)$ corresponds to the initial (upper) state of the transition and $\Psi_{a}(v',R)$ corresponds to the final (lower) state.
The calculated $F (v,v')$ are shown in Fig. \ref{fig:CondonFactors}.
They represent two large groups of partially overlapping vibronic transitions centered at 590 and 620 nm.

In order to obtain the transition dipole moments for the four bands, we assume that the electronic eigenstates of Ba$^{+\ast}$He exactly correspond to the eigenstates of Ba$^{+}$ in the $|L,S,J,M_{J}\rangle$ basis.
We neglect any possible dependence of the transition probability on the internuclear distance.
In fact, at small internuclear separations, the Ba$^{+}$ - He interaction may become stronger than the spin-orbit interaction and lead to the spin-orbit uncoupling, as in the case of alkali-He exciplexes \cite{MoroshkinJCP2006,HoferPRA2006}.
In that case, the electronic states of the excimer correspond to the spin-orbit uncoupled  $|L,M_{L},S,M_{S}\rangle$ states of the metal atom (ion).
We have verified, that due to a stronger spin-orbit interaction in Ba$^{+}$, the spin-orbit coupling is preserved and $J$ remains a good quantum number at the relevant internuclear distances.
The correspondence between the quasimolecular $^{2}\Lambda_{J}$ states and $|L,S,J,M_{J}\rangle$ states is given in Table \ref{eq:TransitionDipole}.
All quasimolecular eigenstates are doubly-degenerate.

\begin{table}
	\centering
	\begin{tabular}{c|c|c}
		$^{2}\Lambda_{\Lambda+\Sigma}$ & $|L,S,J,M_{J}\rangle$ & $D_{3/2,\pm3/2,J',M'_{J}}$ \\\hline\hline
		$A ^{2}\Pi_{1/2}$ & $|2,\dfrac{1}{2},\dfrac{3}{2},\pm\dfrac{3}{2} \rangle$ & $\dfrac{1}{5\sqrt{2}}$  \\
		$B ^{2}\Sigma_{1/2}$ & $|2,\dfrac{1}{2},\dfrac{3}{2},\pm \dfrac{1}{2} \rangle$ & $\dfrac{1}{5\sqrt{3}}$ \\
		$C ^{2}\Pi_{3/2}$ & $|2,\dfrac{1}{2},\dfrac{5}{2},\pm\dfrac{3}{2} \rangle$ & $\dfrac{\sqrt{2}}{5}$ \\
		$D ^{2}\Delta_{5/2}$ & $|2,\dfrac{1}{2},\dfrac{5}{2},\pm\dfrac{5}{2} \rangle$ & $\dfrac{1}{\sqrt{5}}$ \\
		$E ^{2}\Sigma_{1/2}$ & $|2,\dfrac{1}{2},\dfrac{5}{2},\pm\dfrac{1}{2} \rangle$ & $\dfrac{1}{\sqrt{5}}$ \\
		$G ^{2}\Pi_{3/2}$ & $|1,\dfrac{1}{2},\dfrac{3}{2},\pm\dfrac{3}{2} \rangle$ & --  \\
	\end{tabular}
	\caption{Transition dipole moments of Ba$^{+\ast}$He in the units of the reduced dipole moment $(1||D||2)$.}\label{table:DipoleMoments}
\end{table}

Matrix elements of the transition dipole moment, $D_{J,M_{J},J',M'_{J}} = \langle L,S,J,M_{J}|D|L',S',J',M'_{J}\rangle$ are expressed via the reduced transition dipole moment $(L||D||L')$ that is common for all four bands.
Here, $L$ = 1, $L'$ = 2 are the orbital angular momenta of the upper and lower state, respectively, $S$ = $S'$ = 1/2 is the spin, and $J$ = 3/2, $J'$ = 3/2, 5/2 are the total electronic momenta.
By applying Wigner-Eckhart theorem and angular momentum decoupling rules \cite{SobelmanBook} we obtain:
\begin{align}
	\langle L,S,J,M_{J}|D|L',S',J',M'_{J}\rangle \propto (L||D||L') \times \nonumber\\  \times \sqrt{(2J+1)(2J'+1)} \left( \begin{array}{ccc}J & 1 & J'\\-M_{J} & q & M'_{J}\end{array}\right) \left\lbrace \begin{array}{ccc} L & J & S\\J' & L' & 1\end{array} \right\rbrace   \label{eq:TransitionDipole}
\end{align}
Where the round (curly) brackets denote the $3j$ ($6j$) symbols.
Index $q$ equals 0, or $\pm$1 (only the terms with $M_{J} - M'_{J} = q$ give nonzero contributions).
The resulting values of $D_{J,M_{J},J',M'_{J}}$ in the units of the reduced dipole moment $(L||D||L')$ are given in the Table \ref{eq:TransitionDipole}.
According to this calculation, the integrated intensity of the bands $G ^{2}\Pi_{3/2}$ - $A ^{2}\Pi_{1/2}$ and $G ^{2}\Pi_{3/2}$ - $B ^{2}\Sigma_{1/2}$ is 9 times weaker than that of the $G ^{2}\Pi_{3/2}$ - $C ^{2}\Pi_{3/2}$ and $G ^{2}\Pi_{3/2}$ - $D ^{2}\Delta_{5/2}$ bands.

The relative intensities of the peaks in the emission spectra are given by
\begin{equation}
I_{rel} \propto F(v,v') D_{J,M_{J},J',M'_{J}}^{2} N(v) \label{eq:RelativeIntensity} .
\end{equation}
Here, $N(v)$ denotes the relative populations of the vibration levels in the $G ^{2}\Pi_{3/2}$ potential well.

We have also solved the Schr\"{o}dinger
equation for the unbound ground $X
^{2}\Sigma_{1/2}$ state and obtained the corresponding wavefunction $\Psi_{g}(E,R)$.
We then computed the
Franck-Condon factors for the transitions from the five bound states
of the $G ^{2}\Pi_{3/2}$ manifold to this state.
We assume that the transition dipole moment for those transitions is equal to that of the $6 ^{2}P_{3/2} \rightarrow 6 ^{2}S_{1/2}$ transition in a free Ba$^{+}$ ion and does not depend on the internuclear separation.

\section{Discussion \label{Seq:discussion}}

At the temperature of our experiments, the equilibrium (thermalized)
populations of all vibration states except the lowest state should be
very close to zero.
However, it is not clear whether the relatively short lifetime of
the laser-excited $G ^{2}\Pi_{3/2}$ state, $\tau = A^{-1} \approx
10$ ns, is sufficient for thermalization.

\begin{figure} [tbp]
    \includegraphics [width=\columnwidth] {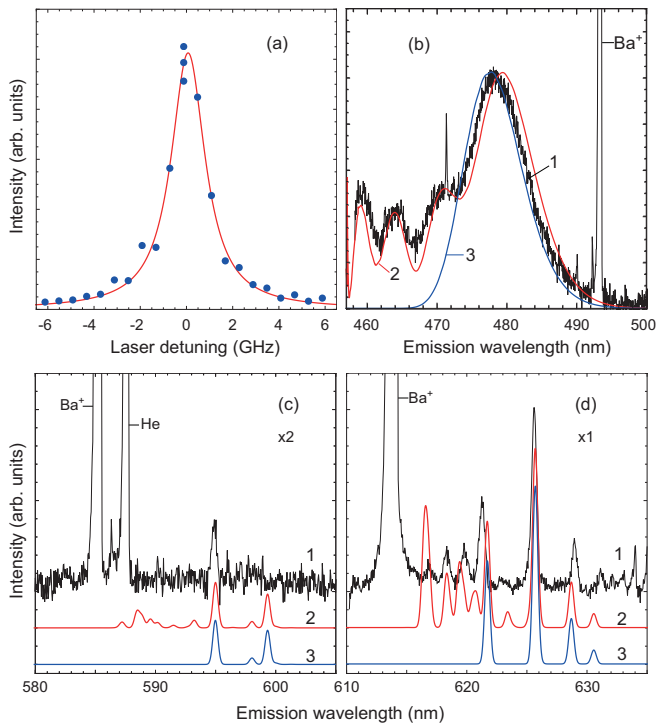}
    \caption{(a) Excitation spectrum of the Ba$^{+}$ 6$^{2}S_{1/2} \rightarrow 6^{2}P_{3/2}$ transition. $T$=1.37 K, $p_{He}$=2.4 mbar. Zero detuning corresponds to transition wavelength in vacuum: $\lambda$ = 454.4 nm. (b)--(d) Emission spectra of the Ba$^{+\ast}$He excimer. $T$=1.57 K, $p_{He}$=6.6 mbar. (b) $G ^{2}\Pi_{3/2} \rightarrow X ^{2}\Sigma_{1/2}$ bound-free transition;  (c) $G ^{2}\Pi_{3/2} \rightarrow A ^{2}\Pi_{1/2}$ and $G ^{2}\Pi_{3/2} \rightarrow B ^{2}\Sigma_{1/2}$ bound-bound transitions; (d) $G ^{2}\Pi_{3/2} \rightarrow C ^{2}\Pi_{3/2}$ and $G ^{2}\Pi_{3/2} \rightarrow D ^{2}\Delta_{5/2}$ bound-bound transitions. Curve 1 (black): experimental data, curve 2 (red): calculated spectrum fitted to the experimental data (see the text), curve 3 (blue): spectrum calculated assuming only $v$ = 0 state populated.}
    \label{fig:SpectraTheorExp}
\end{figure}

In Fig. \ref{fig:SpectraTheorExp}(b) we show the high-resolution experimental
spectrum of the $G ^{2}\Pi_{3/2} \rightarrow X ^{2}\Sigma_{1/2}$
bound-free transition together with the results of the calculation,
assuming that only the $v$ = 0 state is populated.
The main peak centered at 478 nm is reproduced by the calculation reasonably well considering that there are no free parameters except for and only arbitrary vertical scaling.
However, the smaller peaks at shorter wavelengths are not reproduced at all.
We therefore assume incomplete thermalization of the vibronic states and attempt to fit the experimental spectrum by adjusting the populations of these five states.
The result of this fit is shown in the same figure, and it closely reproduces all four peaks.
The resulting relative populations $N(v)$ are equal to 31\%, 10\%, 13\%, 11\%, and 35\% for $v$ = 0, 1, 2, 3, and 4, respectively.

In Figs. \ref{fig:SpectraTheorExp}(c), (d) we show high-resolution experimental and calculated spectra of the bound-bound transitions.
Each vibronic transition is represented by a Gaussian curve with 0.3
nm spectral width, which approximately corresponds to the resolution
of our spectrometer.
The height of each peak is determined using Eq. \ref{eq:RelativeIntensity}.
We present two calculated results.
The first calculation assumes $N(v=0)$ = 100\%, $N(v\neq0)$ = 0, \textit{i.e.}, complete thermalization.
The second calculation uses the same relative populations $N(v)$ as those obtained from the fitting of the bound-free $G ^{2}\Pi_{3/2} \rightarrow X ^{2}\Sigma_{1/2}$ band.
In both cases, we must introduce an extra shift of the calculated spectra by 2.0 nm towards longer wavelengths to approximately fit the peak positions.
For the $G ^{2}\Pi_{3/2} \rightarrow C ^{2}\Pi_{3/2}$ and $G
^{2}\Pi_{3/2} \rightarrow D ^{2}\Delta_{5/2}$ bands, the resulting
computed spectra fit the observed peak positions reasonably well.
A better fit is provided by the calculation using fitted relative
populations, although some peak heights disagree with the
experimental data.
For the $G ^{2}\Pi_{3/2} \rightarrow A ^{2}\Pi_{1/2}$ and $G
^{2}\Pi_{3/2} \rightarrow B ^{2}\Sigma_{1/2}$ bands, the
experimental spectra are too weak for a detailed comparison, and only the strongest spectral component at 595 nm could be distinguished.

We estimated the rotation constants for the bound states of Ba$^{+\ast}$He studied here: $B =\hbar^{2}/2\mu R_{0}^{2}$.
Here, $\mu$ denotes the reduced mass of the molecule and $R_{0}$ is the interatomic distance of the corresponding state.
For all states studied here, $B \simeq$ 0.5 cm$^{-1}$.
The resulting rotational structure could not be resolved in our experiment.

It is very likely that the same type of Ba$^{+\ast}$He quasimolecules can also be obtained in superfluid He, where larger polyatomic complexes can be expected, in analogy with alkali-He exciplexes \cite{HiranoPRA2003,NettelsPRL2005}.
Since the structure of the complex is determined by the shape of the valence electron density distribution of Ba$^{+}$, the states with $L$ = 2 will give rise to qualitatively different structures from those of the $L$ = 1 exciplexes studied earlier \cite{HiranoPRA2003,NettelsPRL2005}.
Some idea of possible exciplex structures can be obtained from the angular distributions of the electron density shown on the left-hand side of Fig. \ref{fig:PotentialDiagram}.
In particular, the states $B$ and $C$ may be able to bind several He atoms on a ring around the waist of the orbital, in a plane orthogonal to the symmetry axis, plus two more He atoms on the axis.

It is also important to note that the states $A$, $B$, $C$, and $D$ are metastable since the transitions from the parent 5$^{2}D_{3/2,5/2}$ states to the ground state in the free ion are forbidden.
It may be possible to accumulate a sufficient population in these states to study the absorption spectra of the same bound-bound transitions by the methods of high-resolution laser spectroscopy.
In that case, the rotational structure could be resolved, which would provide much more detailed information about the structure and energetics of the quasimolecules and their temperature.

We are currently setting up a new experiment with the aim of trapping of Ba$^{+}$ ions in a static electric field in superfluid He \cite{BatulinJLTP2014}.
The formation of metastable quasimolecules in states $A$, $B$, $C$, and $D$ may lead to the depletion of the ionic population in the trap and therefore requires more detailed study.

% Specify following sections are appendices. Use \appendix* if there
% only one appendix.
%\appendix*

\begin{acknowledgments}

We thank M. Mella and F. Cargnoni for sending us the numerical data of the interaction potentials \cite{MellaJPCA2014} and D. Marienko and Y. P. Lee for the critical reading of the manuscript.
We acknowledge fruitful discussions with P. Leiderer.
This work was supported by JSPS KAKENHI grant No. 24000007.

\end{acknowledgments}

%\bibliography{BaHeExcimer}

\end{document}